\documentclass[showpacs, pra, aps, superscriptaddress,twocolumn,10pt]{revtex4}
\usepackage{mathrsfs}
\usepackage{array}
\usepackage{amsmath}
\usepackage{graphicx}
\usepackage{amstext}
\usepackage{amsfonts}

\begin{document}
\title{Dynamics of open quantum systems with initial system-reservoir correlations}
\author{Hua-Tang Tan}
\affiliation{Department of Physics and Center for Quantum
information Science, National Cheng Kung University, Tainan 70101,
Taiwan} \affiliation{Department of Physics, Huazhong Normal
University, Wuhan 430079, China}
\author{Wei-Min Zhang}
\email{wzhang@mail.ncku.edu.tw}
\affiliation{Department of Physics
and Center for Quantum information Science, National Cheng Kung
University, Tainan 70101, Taiwan}
\begin{abstract}
In this paper, the exact dynamics of open quantum systems in the
presence of initial system-reservoir correlations is investigated
for a photonic cavity system coupled to a general non-Markovian
reservoir. The exact time-convolutionless master equation
incorporating with initial system-reservoir correlations is
obtained. The non-Markovian dynamics of the reservoir and the
effects of the initial correlations are embedded into the
time-dependent coefficients in the master equation. We show that the
effects induced by the initial correlations play an important role
in the non-Markovian dynamics of the cavity but they are washed out
in the steady-state limit in the Markovian regime. Moreover, the
initial two-photon correlation between the cavity and the reservoir
can induce nontrivial squeezing dynamics to the cavity field.
\end{abstract}

 \maketitle
\section{Introduction}

The study of dynamics of open quantum systems continuously receives
attentions because of its fundamental importance in quantum physics
and also because of the rapid development of quantum technologies.
Previous studies on the dynamics of open quantum systems mainly lie on
the Lindblad-type master equation \cite{bm1,bm2,bm3}, where the
characteristic time of the environment is sufficiently shorter
than that of the system such that the non-Markovian memory
effect is negligible, and so does for initial system-reservoir
correlations. However, the new development in ultrafast photonics,
ultracold atomic physics, nanoscience and technology as well as
quantum information science strongly suggests that the non-Markovian
dynamics in ultrafast and ultrasmall open systems should play an
important role, and the associated effects (including the initial
system-reservoir correlations) should be fully taken into account.
To this end, the more rigorous approach is demanded for the study of
non-Markovian dynamics of open quantum systems incorporating with the
initial system-reservoir correlations.

The exact description of open quantum systems has indeed been
explored extensively in the literature, mainly focusing on quantum
Brown motion based on Feynman-Vernon influence functional
\cite{Fey63118,Cal83587,Haa852462,Hu922843,Hal962012,Food01105020}
and stochastic diffusion Schr\"{o}dinger equation
\cite{Str981699,Str994909,Yu04062107}. Extending the Feynman-Vernon
influence functional to other open quantum systems has also achieved
a great success recently, including the exact master equation for
electron systems and the nonequilibrium quantum transport theory in
various nanostructures \cite{Tu08235311,Tu09631,Jin10083013} and the
exact master equation for micro- or nanocavities in photonic
crystals and the quantum transport theory for photonic crystals
\cite{Xio10012105,Wu1018407,Lei104570}. However, in most of these
investigations, the system and the reservoir are often assumed to be
initially uncorrelated with each other \cite{Leg871}. Realistically,
it is possible and often unavoidable in experiments that the system
and its environment are correlated closely at the beginning,
especially for the cases of the system strongly coupled to the
reservoir \cite{ee}. Various initial-correlation induced effects
have been investigated in different open quantum systems \cite{src0,
src1, src2, src3, src4, src5,src7, src8,src6,src9, src10}. For
example, it has been recently shown that the initial correlations
between a qubit and its environment can lead to the distance growth
of two quantum states over its initial value \cite{src7, src8}. It
has also been demonstrated that the initial correlations have
nontrivial differences in quantum tomography process \cite{src6}.
Besides, it has been found that the initial system-reservoir
correlations have significant effects on the entanglement in a
two-qubit system \cite{src9, src10}.

In this paper, the dynamics of open quantum systems in the presence
of initial system-reservoir correlations is investigated with a
photonic cavity system coupled to a non-Markovian reservoir as a
specific example. By solving the exact dynamics of the cavity
system, the effects of the initial correlations are explicitly built
into the equations of motion for the intensity and the two-photon
correlation function of the cavity field. We then obtain the exact
master equation incorporating with the initial correlations which
induce new terms and also modify the time-dependent dissipation and
fluctuation coefficients in the master equation. Taking a nanocavity
coupled to a coupled resonator optical waveguide (serving as a
structured reservoir) as an experimentally realizable system, we
find that the effects of the initial correlations are fragile for a
Markovian reservoir but play an important role in the non-Markovian
regime. In fact, in the strong non-Markovian regime, the initial
two-photon correlation between the cavity and the reservoir can
induce oscillating squeezing dynamics in the cavity. But in the
Markovian regime, the initial correlations will be washed out in the
steady-state limit.

The rest of the paper is organized as follows. In Sec.~II, the
dynamics of open quantum systems with initial system-reservoir
correlations is formulated for a photonic cavity system coupled to a
general non-Markovian reservoir. In Sec.~III, we construct the exact
time-convolutionless master equation incorporating with the initial
correlations, where the effects from the initial correlations are
explicitly embedded into the time-dependent coefficients in the
master equation. In Sec.~IV, an experimentally realizable example is
considered to analytically and numerically examine the influence of
the initial correlations on the dynamics of open quantum systems.
At last, a summary is given in Sec. V.

\section{Non-Markovian dynamics with initial system-reservoir correlations}
To be specific, we consider here a single-mode photonic cavity
system coupled to a general non-Markovian reservoir, where the
single-mode cavity system could be a nanocavity in nanostructures or
photonic crystals, and the non-Markovian environment may be a
structured photonic reservoir \cite{stru-reservoir}. The Hamiltonian
of the system can be expressed as a Fano-type model of a localized
state coupled with a continuum \cite{Fano611866}:
\begin{eqnarray}
H=\omega_c a^\dag a+\sum_{k}\omega_k b_k^\dag b_k\ +\sum_kV_k (a
b_k^\dag +b_k a^\dag),\label{H1}
\end{eqnarray}
where the first term is the Hamiltonian of the cavity field with
frequency $\omega_c$, and $a^\dag$ and $a$ are the creation and
annihilation operators of the cavity field; the second term
describes a general non-Markovian reservoir which is modeled as a
collection of infinite photonic modes, where $b_k^\dag$ and $b_k$
are the corresponding creation and annihilation operators of the
$k$-th photonic mode with frequency $\omega_k$. The third term
characterizes the system-reservoir coupling with the coupling
strength $V_k$ between the cavity field and the $k$-th photonic
mode. For convenience, we take $\hbar=1$ throughout the paper.

We shall use the equation of motion approach to solve the dynamics
of the cavity system and the reservoir, from which the
general initial correlations between the cavity and the reservoir can
be fully taken into account. The time evolution of the cavity field
operator $a(t)=e^{iHt}ae^{-iHt}$ and the reservoir field operators
$b_k(t)=e^{iHt}b_ke^{-iHt}$ in the Heisenberg picture obey the
equations of motion
\begin{subequations}
\begin{align}
&\frac{d}{dt}a(t)=-i[a(t), H]=-i\omega_c a(t)-i\sum_k V_k b_k(t),\\
&\frac{d}{dt}b_k(t)=-i[b_k(t),H]=-i\omega_k b_k(t)-iV_k a(t).
\label{bk}
\end{align}
\end{subequations}
Solving Eq.~(\ref{bk}) for $b_k(t)$
\begin{eqnarray}
b_k(t)=b_k(0)e^{-i\omega_k t}-iV_k\int_0^t d\tau
a(\tau)e^{-i\omega_k (t-\tau)},
\end{eqnarray}
we obtain
\begin{align}
\frac{d}{dt}a(t)=-i\omega_c a(t) -\int_0^t d\tau g(t-\tau)a(\tau)
\nonumber\\-i\sum_k V_k b_k(0)e^{-i\omega_k t}. \label{lat}
\end{align}
Here, the memory kernel $g(\tau)=\sum_{k}|V_k|^2e^{-i\omega_k\tau}$
characterizes the non-Markovian dynamics of the reservoir. For a
continuous reservoir spectrum, we have $g(\tau)=\int_{0}^\infty
\frac{d\omega}{2\pi}J(\omega)e^{-i\omega\tau}$, where $J(\omega)=2\pi
\varrho(\omega)|V(\omega)|^2$ is
the spectral density of the reservoir, with $\varrho(\omega)$ being the
density of states and $V(\omega)$ the coupling between the cavity
and the reservoir in the frequency domain.

Because of the linearity of Eq.(\ref{lat}), the cavity field
operator $a(t)$ can be expressed, in terms of the initial field
operators $a(0)$ and $b_k(0)$ of the cavity and the reservoir, as
\begin{align}
a(t)=u(t)a(0)+f(t) , \label{at}
\end{align}
where the time-dependent coefficient $u(t)$ and $f(t)$ are
determined from Eq.(\ref{lat}) and given by
\begin{subequations}
\begin{align}
\frac{d}{dt}u(t)=-i\omega_c u(t)-\int_0^t d\tau
g(t-\tau)u(\tau),\label{ut}
\\
\frac{d}{dt}f(t)= -i\omega_c f(t)-\int_0^t d\tau g(t-\tau)f(\tau) \label{ft} \nonumber\\
 -i\sum_kV_k b_k(0)e^{-i\omega_k \tau},
\end{align}
\end{subequations}
subjected to the initial conditions $u(0)=1$ and $f(0)=0$. The
integrodifferential equation (\ref{ut}) shows that $u(t)$ is just
the propagating function of the cavity field (the retarded Green
function in nonequilibrium Green function theory \cite{green}). In
addition, $f(t)$ is in fact an operator coefficient and its solution
can be obtained analytically from the inhomogeneous equation of
Eq.~(\ref{ft}):
\begin{align}
f(t)=-i\sum_kV_k b_k(0)\int_0^t d\tau e^{-i\omega_k \tau}u(t-\tau).
\label{ress}
\end{align}

From Eqs.~(\ref{at})-(\ref{ress}) we can determine the exact non-Markovian
dynamics of the cavity field coupled to a general reservoir with
arbitrary initial system-reservoir correlations, upon a given initial
state $\rho_{\rm tot}(0)$ of the whole system. In the Heisenberg
picture, quantum states are time-independent. Once $\rho_{\rm
tot}(0)=\rho_{\rm tot}$ is given, the time evolution of any physical
observable can be obtained directly from
Eqs.~(\ref{at})-(\ref{ress}) through the relation
\begin{align}
\langle f(a^\dag(t),a(t))\rangle = {\rm
tr}[f(a^\dag(t),a(t))\rho_{\rm tot}].
\end{align}
For example, the time evolution of the expectation values $\langle
a(t) \rangle$, $n(t) \equiv \langle a^\dag(t) a(t) \rangle$, and
$s(t) \equiv \langle a(t) a(t)\rangle$, which respectively describe
the cavity amplitude, the cavity intensity, and the two-photon
correlation of the cavity field, can be expressed explicitly by the
following solution
\begin{subequations}
\label{e}
\begin{align}
&\langle a (t) \rangle=u(t)\langle a (0)\rangle + \upsilon_0(t),\label{e1}\\
& n(t)=|u(t)|^2  n(0)
+2{\rm Re}[u^*(t)\nu_1(t)]+\upsilon_1(t),\label{e2}\\
& s(t)=u^2(t) s (0) +2u(t)\nu_2(t)+\upsilon_2(t) , \label{e3}
\end{align}
\end{subequations}
where $\langle a (0)\rangle, n(0)$ and $ s (0)$ are the
corresponding initial conditions. Other time-dependent functions in
Eq.~(\ref{e}) are given by
\begin{subequations}
\label{core}
\begin{widetext}
\begin{align}
\nu_1(t)&=\langle a^\dag(0)f(t)\rangle
=-i\int_0^t\sum_k V_k \langle a^\dag(0)b_k(0)\rangle e^{-i\omega_k \tau}u(t-\tau)d\tau,\label{u1}\\
\nu_2(t)&=\langle a(0)f(t)\rangle
=-i\int_0^t\sum_k V_k \langle a(0) b_k(0)\rangle e^{-i\omega_k \tau}u(t-\tau)d\tau,\label{u2}\\
\upsilon_0(t)&=-i\int_0^t\sum_k V_k \langle b_k(0)\rangle
e^{-i\omega_k\tau}u(t-\tau)d\tau,\label{u0}\\
\upsilon_1(t)&=\langle f^\dag(t)f(t)\rangle =\int_0^t d\tau\int_0^t
d\tau'\sum_{kk'} V^*_k V_{k'}\langle b^\dag_k(0)
b_{k'}(0)\rangle e^{-i(\omega_{k'}\tau'-\omega_k\tau)}u^*(t-\tau)u(t-\tau'),\label{v1}\\
\upsilon_2(t)&=\langle f(t)f(t)\rangle =-\int_0^t d\tau\int_0^t
d\tau'\sum_{kk'} V_k V_{k'}\langle b_k(0) b_{k'}(0)\rangle
e^{-i(\omega_k\tau+\omega_{k'}\tau')}u(t-\tau)u(t-\tau')\label{v2}.
\end{align}
\end{widetext}
\end{subequations}
In these solutions, $\upsilon_j(t)$ ($j=0,1,2$) characterize
respectively the contributions from the initial field amplitudes
$\langle b_k(0)\rangle$, the initial photon scattering amplitudes
$\langle b^\dag_k(0) b_{k'}(0)\rangle$ and the initial two-photon
correlations $\langle b_k(0) b_{k'}(0)\rangle$ of all the photonic
modes in the reservoir. While $\nu_1(t)$ and $\nu_2(t)$ correspond
to the contributions from the different initial system-reservoir
correlations $\langle a(0)b_k^\dag(0)\rangle$ and $\langle
a(0)b_k(0)\rangle$, respectively.

If the initial state of the total system is uncorrelated, and the
reservoir is in a thermal equilibrium state, i.e.,
\begin{align}
\rho_{\rm tot}(0)=\rho(0) \times \rho_R(0), ~~
\rho_R(0)=\frac{e^{-\beta H_R}}{tr e^{-\beta H_R}},
\end{align}
with $H_R=\sum_k\omega_kb_k^\dag b_k$ and $\beta=1/k_BT$ being the initial temperature
of the reservoir, it is easy to check that $\nu_i(t)=0, \upsilon_i(t)=0$ except for
$v_1(t)$ which is given by
\begin{align}
\upsilon_1(t) =\int_0^t d\tau\int_0^t d\tau'
u(t-\tau')\widetilde{g}(\tau'-\tau) u^*(t-\tau). \label{vts}
\end{align}
Here, $\widetilde{g}(\tau)=\sum_k |V_k|^2 \bar{n}_k e^{-i\omega_k
\tau}$ and $\bar{n}_k=\langle b^\dag_k(0) b_{k}(0)\rangle=
1/(e^{\beta \omega_k} -1)$ is the initial photonic distribution
function of the reservoir. Then Eq.~(\ref{e}) reproduces the same
solution solved from the exact master equation without initial
system-reservoir correlations \cite{Xio10012105}. However, as we
see, the exact non-Markovian dynamics in Eq.(\ref{e}) derived via
the equation of motion approach has explicitly included the effects
induced by the initial correlations between the system and the
reservoir.

\section{Exact master equation with initial system-reservoir correlations}
To further understand the effects of the initial system-reservoir
correlations on the dynamics of open quantum systems, we shall
attempt to derive the exact master equation for the reduced density
matrix of the cavity system $\rho(t)$. In the literature, exact
master equations for open systems are mostly derived without initial
correlations, such as the systems associated with quantum Brown
motions \cite{Hu922843,Hal962012,Food01105020}, quantum dot systems
in various nanostructures \cite{Tu08235311,Tu09631} and cavity
systems coupled to structured reservoirs as well as general
non-Markovian reservoirs \cite{An07042127,Xio10012105,Wu1018407}.
Here, we concentrate the exact master equation for the photonic
system in the presence of initial Gaussian correlated states. Based
on the bilinear operator structure of the system as well as the
techniques in deriving exact master equation for the cavity system
described by Eq.~(\ref{H1}) \cite{Xio10012105,Wu1018407}, the master
equation with the initial system-reservoir correlations would have a
general time-convolutionless form as follows:
\begin{align}
\dot{\rho}(t)=& -i\Delta(t)[a^\dag a, \rho] \notag \\
& +\gamma_1(t)(2a\rho a^\dag -a^\dag a\rho-\rho_a a^\dag a)\nonumber\\
&+\gamma_2(t)(a\rho a^\dag+a^\dag\rho a -a^\dag a\rho-\rho a a^\dag)\nonumber\\
&+\gamma_3^*(t)(2a\rho a-aa\rho-\rho aa)\nonumber\\
&+\gamma_3(t)(2a^\dag \rho a^\dag -a^\dag a^\dag \rho-\rho a^\dag
a^\dag),\label{me}
\end{align}
where the coefficient $\Delta(t)$ is the renormalized cavity
frequency, $\gamma_1(t)$ and $\gamma_2(t)$ usually denote
respectively the dissipation (damping) and fluctuation (noise) due
to the back-reactions between the system and the reservoir, and
$\gamma_3(t)$ is related to a two-photon decoherence process. As we
see, the first three terms have the standard form as the exact
master equation for the Hamiltonian in Eq.~(\ref{H1}) without the
initial correlations \cite{Xio10012105,Wu1018407}, but with the
coefficients modified by the initial correlation $\langle
a(0)b^\dag_k(0) \rangle$. The last two terms are contributed from
the two-photon correlation $\langle b_k(0)b_{k'}(0) \rangle$ in the
reservoir as well as by the initial system-reservoir two-photon
correlation $\langle a(0)b_k(0) \rangle$.

To figure out the time-convolutionless coefficients in
Eq.~(\ref{me}), we compute the physical observables in Eq.~(\ref{e})
from the above master equation. From Eq.~(\ref{me}), it is easy to
find that
\begin{subequations}
\label{opem}
\begin{align}
&\frac{d}{dt}\langle a(t)\rangle=-[\gamma_1(t)+i\Delta(t)]\langle a(t) \rangle, \\
&\frac{d}{dt}n(t)=-2\gamma_1(t)n(t)+2\gamma_2(t), \\
&\frac{d}{dt}s(t)=-2[\gamma_1(t)+i\Delta(t)]s(t)-2\gamma_3(t).\label{meq}
\end{align}
\end{subequations}
On the other hand, with Eq.(\ref{at}) we obtain
\begin{eqnarray}
\frac{d}{dt}a(t)=\frac{\dot{u}(t)}{u(t)}a(t)-\frac{\dot{u}(t)}{u(t)}f(t)
+\dot{f}(t).\label{at3}
\end{eqnarray}
Note that the photonic modes in the reservoir usually cannot be a
coherent state so that $\langle b_k(0)\rangle=0$. Then using
Eq.~(\ref{at3}), we find that
\begin{subequations}
\label{opem1}
\begin{align}
&\frac{d}{dt}\langle a(t)\rangle=\frac{\dot{u}(t)}{u(t)}\langle a(t)\rangle, \\
&\frac{d}{dt}n(t)=2{\rm
Re}\Big[\frac{\dot{u}(t)}{u(t)}\Big]n(t) +\dot{\upsilon}_1(t) -2{\rm Re}\Big[\frac{\dot{u}(t)}{u(t)}\Big]\upsilon_1(t)\nonumber\\
&~~~~~~~~~~ ~~~~~~ +2{\rm Re}\Big[u^*(t)\dot{\nu}_1(t)-\frac{\dot{u}(t)u^*(t)}{u(t)}\nu_1(t)\Big] ,\\
&\frac{d}{dt}s(t)=2\frac{\dot{u}(t)}{u(t)}s(t) +\dot{\upsilon}_2(t) -2\frac{\dot{u}(t)}{u(t)}\upsilon_2(t)
 \nonumber\\
&~~~~~~~~~~ ~~~~~~~~~~~~~~~  +2u(t)\dot{\nu}_2(t)
-2\dot{u}(t)\nu_2(t).\label{ap1}
\end{align}
\end{subequations}
By comparing Eq.~(\ref{opem}) with Eq.~(\ref{opem1}), the
coefficients $\Delta(t)$ and $\gamma_j(t)$ in the master equation
can be uniquely determined and given by
\begin{subequations}
\label{ecoff}
\begin{align}
&\Delta(t)=-{\rm Im}\Big[\frac{\dot{u}(t)}{u(t)}\Big],~~
\gamma_1(t)=-{\rm Re}\Big[\frac{\dot{u}(t)}{u(t)}\Big],\\
&\gamma_2(t)=\dot{\upsilon}_1(t)+2{\rm Re}\Big[ u(t)\dot{\nu}_1^*(t)
-\frac{\dot{u}(t)}{u(t)}[ \upsilon_1(t)+u^*(t)\nu_1(t)]\Big],\\
&\gamma_3(t)=-\frac{1}{2}\dot{\upsilon}_2(t)
+\frac{\dot{u}(t)}{u(t)}\upsilon_2(t)-
u(t)\dot{\nu}_2(t)+\dot{u}(t)\nu_2(t),
\end{align}
\end{subequations}
which shows that the coefficients $\gamma_2(t)$ and $\gamma_3(t)$ in
the master equation depend explicitly on the initial correlations
$\langle a(0)b_k^\dag(0)\rangle$ and $\langle a(0)b_k(0)\rangle$ in
the presence of the initial Guassian correlated states of the whole
system.

If the reservoir is initially in a thermal state uncorrelated to the
system, we have $\langle a(0)b^\dag_k(0) \rangle =\langle a(0)b_k(0)
\rangle=\langle b_k(0)b_{k'}(0) \rangle = 0$ except for $\langle
b^\dag_k(0)b_{k'}(0)=\bar{n}_k$. Accordingly, from Eq.(\ref{core}) we have
$\upsilon_2(t)=0=\nu_1(t)=\nu_2(t)$ so that $\gamma_3(t)=0$ and
\begin{eqnarray}
\gamma_2(t)=\dot{\upsilon}_1(t)-2{\rm
Re}\Big[\frac{\dot{u}(t)}{u(t)}\Big]\upsilon_1(t),
\end{eqnarray}
where $\upsilon_1(t)$ is then given by Eq.~(\ref{vts}).
Consequently, the master equation (\ref{me}) in this situation is
reduced to the exact master equation for the cavity system coupled
with a general non-Markovian reservoir presented recently in
Ref.~\cite{Xio10012105,Wu1018407}, which is obtained originally
using the Feynman-Vernon influence functional. In addition, if there
are no initial correlations but the reservoir involves initially
two-photon correlation, namely, $\langle a(0)b^\dag_k(0) \rangle
=\langle a(0)b_k(0) \rangle=0$ but $\langle b_k(0)b_{k'}(0) \rangle
\neq 0$, then we have $\nu_1(t)=0=\nu_2(t)$ but
$\upsilon_2(t)\neq0$. As a result, the coefficient $\gamma_3(t) \neq
0$, which induces a two-photon decoherence process in the cavity
\cite{TMSS1}. However, if the initial states of the whole system
only contains the two-photon correlation $\langle a(0)b_k(0)\rangle$
but the reservoir itself stays in an initial thermal state, then we
have $\nu_1(t)=0=\upsilon_2(t)$ but $\nu_2(t) \neq 0$. This
situation also leads to a non-zero $\gamma_3(t)$ which is
essentially equivalent to the situation in which the reservoir
involves initially two-photon correlation but without the initial
system-reservoir correlations.

Therefore, the master equation, Eq.~(\ref{me}) with the
time-dependent coefficients in Eq.~(\ref{ecoff}), describes the
exact non-Markovian dynamics of a cavity system coupled with a
general reservoir involving two-photon correlation in the presence
of the quadratic-type initial correlations between the system and
reservoir. It shows explicitly that the initial correlation $\langle
a(0)b_k^\dag(0)\rangle$ modifies the fluctuation coefficient
$\gamma_2(t)$ but without altering the damping (dissipation) rate
$\gamma_1(t)$, which in turn changes the cavity field intensity
given by Eq.~(\ref{e2}) without changing the cavity field amplitude
of Eq.~(\ref{e1}). The initial correlation $\langle
a(0)b_k(0)\rangle$ affects on the two-photon decoherence process
which leads to a two-photon process $s(t)=\langle a(t)a(t) \rangle$
of the cavity field. It should be pointed out that if the system and
the reservoir are initially in non-Gaussian correlated states, the
form of Eq.~(\ref{me}) may need to be modified further.
Nevertheless, the master equation of Eq.~(\ref{me}) is exact for the
initial Gaussian
 correlated states of the whole system, and
it remains in a time-convolutionless form in which the non-Markovain
memory dynamics is fully embedded into the time-dependent
coefficients. As we see, all these time-dependent coefficients are
determined by a unique function, the cavity field propagating
function $u(t)$, through the relations given by Eqs.~(\ref{ecoff})
and (\ref{core}). While the propagating function $u(t)$ is
determined by Eq.~(\ref{ut}) in which the integral kernel contains
all the non-Markovian memory effects characterizing the
back-reactions between the system and the reservoir.

\section{Examples with initial system-reservoir correlations}

In this section, we shall take two different initial correlated
states to examine the effects of the initial correlation on the
non-Markovian dynamics in such an open system. To be more specific,
we consider an experimentally realizable nanocavity system. Fig.~1
is a schematic plot for a single-mode nanocavity coupled to a
coupled resonator optical waveguide (CROW) structure. The nanocavity
could be a point defect created in photonic crystals and the
waveguide consists of a linear defects in which light propagates due
to the coupling of the adjacent defects. The CROW is called as a
structured reservoir which possess strong non-Markovian effects
\cite{FanoAnderson-2,Wu1018407}. The Hamiltonian of the whole system
is given by
\begin{eqnarray}
H&=&\omega_ca^\dag a+\sum_{n}\omega_0b_n^\dag b_n
+\lambda (ab_1^\dag+b_1a^\dag)\nonumber\\
&&-\sum_{n}\lambda_0(b_nb_{n+1}^\dag+b_{n+1}b_n^\dag),\label{H2}
\end{eqnarray}
where $a$ and $a^\dag$ are the annihilation and creation operators
of the nanocavity field with frequency $\omega_c$, and the
annihilation and creation operators $b_n$ and $b_n^\dag$ describe
the resonators at site $n$ in the waveguide with identical frequency
$\omega_0$. The frequencies $\omega_c$ and $\omega_0$ are tunable by
changing the size of the relevant defects. The third terms describes
the coupling of the nanocavity field to the resonator at the first
site in the waveguide with the coupling strength $\lambda$ which is
also controllable experimentally by adjusting the distance between
defects. The last term characterizes the photon hopping between two
consecutive resonators in the waveguide structure with the
controllable hopping amplitude $\lambda_0$.
\begin{figure}
\centerline{\scalebox{0.35}{\includegraphics{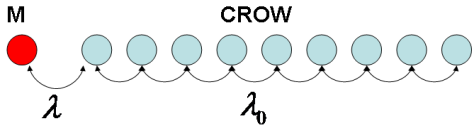}}} \caption{A
schematic plot of a nanocavity (M) coupled to a CROW structure.}
\end{figure}

Consider the waveguide is semi-infinite long, then performing the
following Fourier transformation
$b_k=\sqrt{2/\pi}\sum_{n=1}^\infty\sin(nk)b_n$, where the operators
$b_k$ and $b_k^\dag$ correspond to the Bloch modes of the waveguide,
the Hamiltonian of Eq.~(\ref{H2}) becomes
\begin{eqnarray}
H&=&\omega_ca^\dag a+\sum_{k}\omega_kb_k^\dag b_k
+\sum_{k}g_k(ab_k^\dag+b_ka^\dag),\label{H3}
\end{eqnarray}
where
\begin{align}
\omega_k=\omega_0-2\lambda_0\cos k~,
~~g_k=\sqrt{\frac{2}{\pi}}\lambda\sin k. \label{strength}
\end{align}
with $0\le k\le \pi$.  As we see, Eq.~(\ref{H3}) is reduced to the
same form of Eq.~(\ref{H1}) for the system considered in
Secs.~II-III.

\subsection{Initially system-reservoir correlated squeezed state}
\begin{figure*}
\centerline{\scalebox{0.2}{\includegraphics{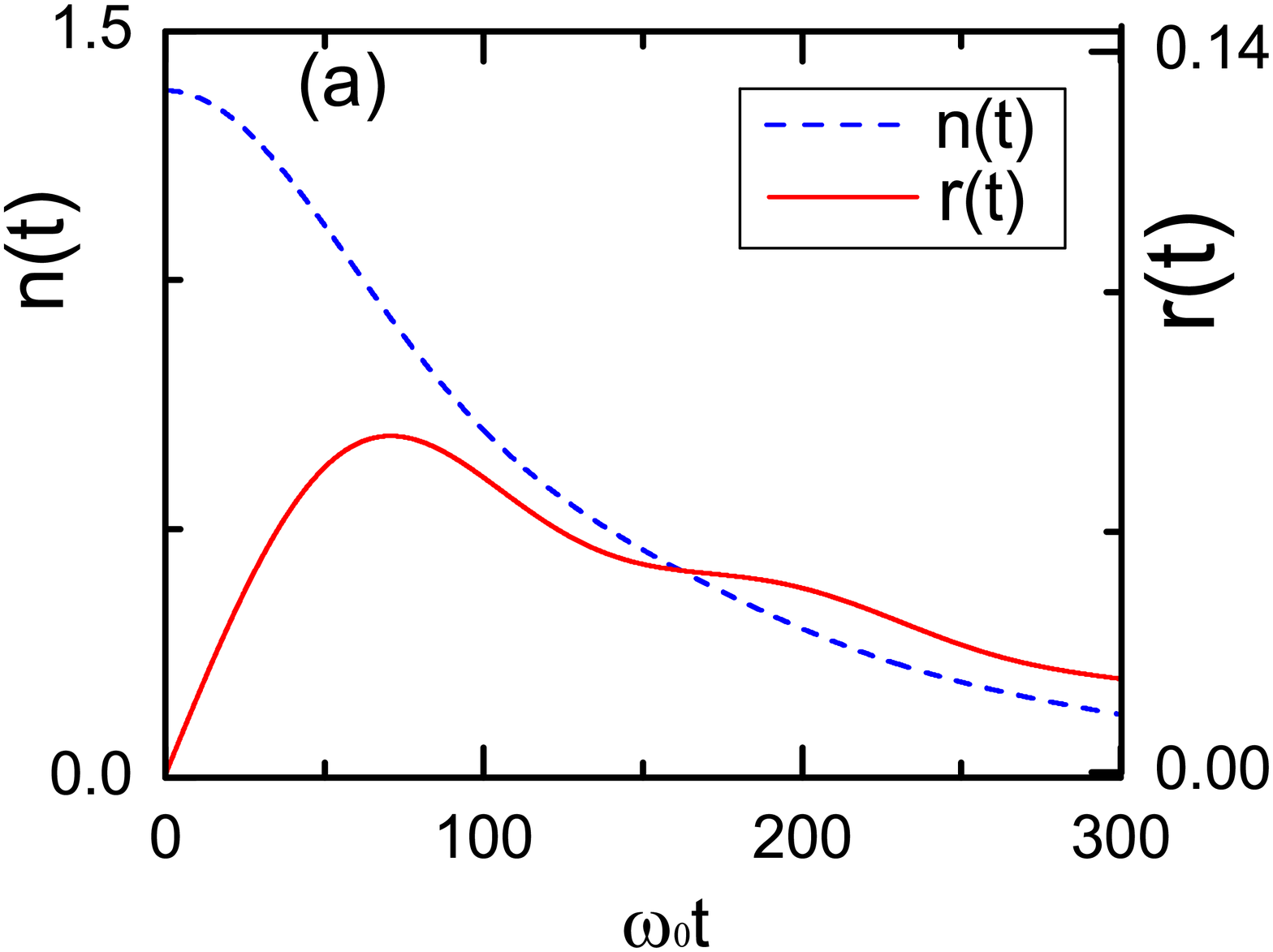}}
\scalebox{0.2}{\includegraphics{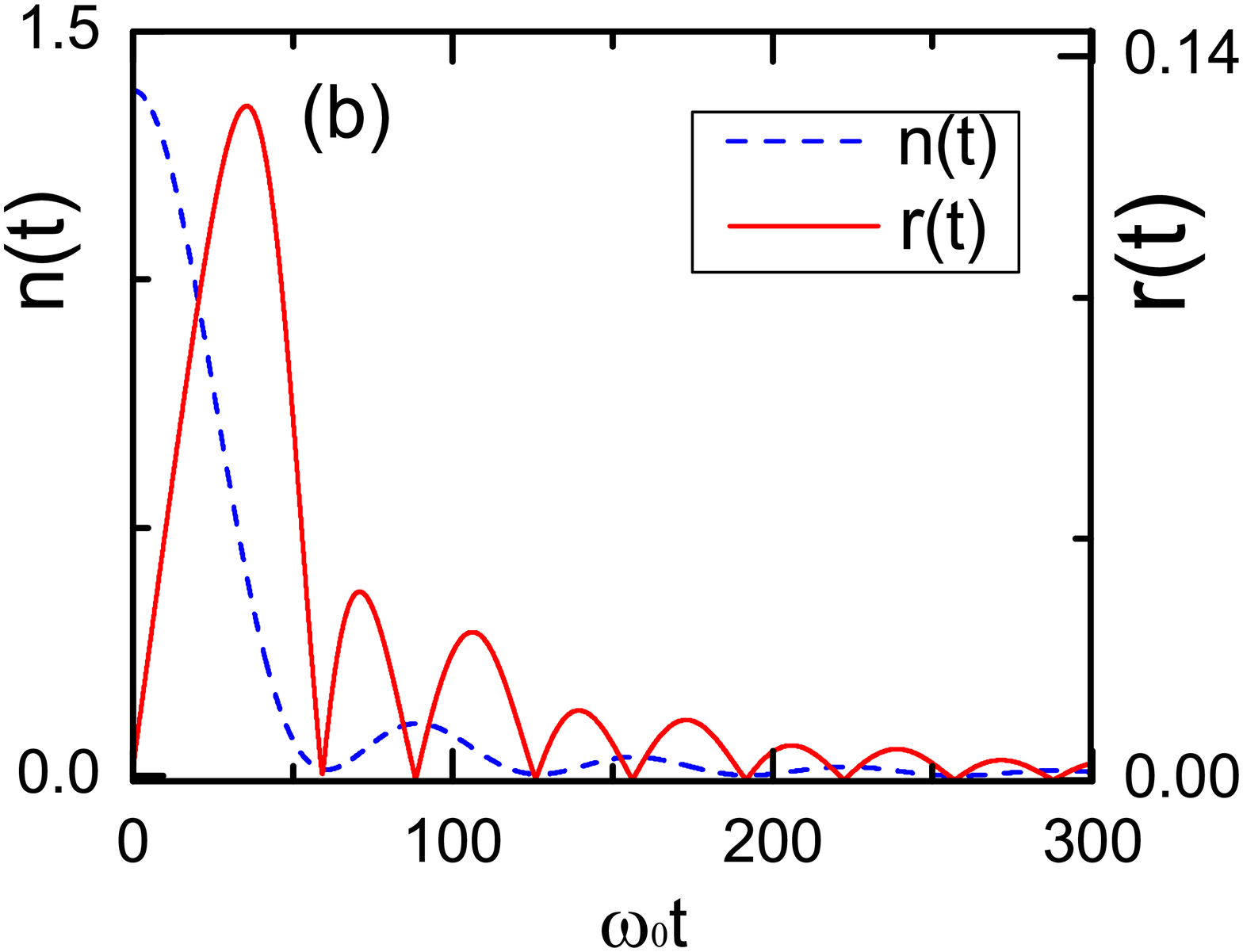}}
\scalebox{0.2}{\includegraphics{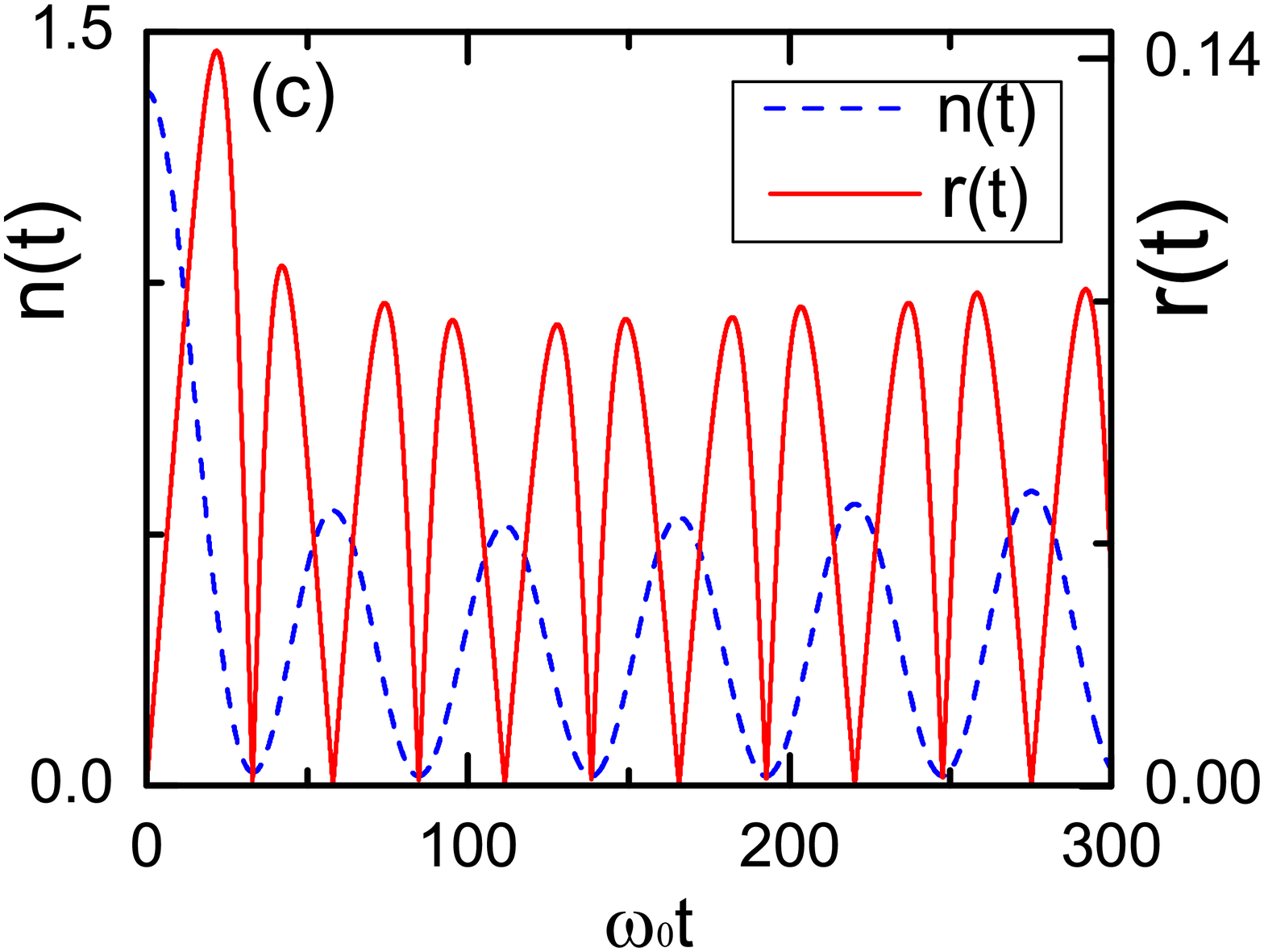}}} \caption{The time evolution
of the cavity field intensity , i.e. the average photon number
$n(t)=\langle a^\dag (t) a(t) \rangle$ and the squeezing parameter
$r(t)$ in the presence of the initial two-photon
correlation between the system and the reservoir. The parameters taken in the plots are
$\omega_c=\omega_0$, $\lambda_0=0.025\omega_0$, $r_s=1.0$,
$\theta_s=0$, with $\eta=0.4$ in (a), $\eta=1.2$ in (b), and
$\eta=2.0$ in (c).} \label{fig2}
\end{figure*}

For the above specific physical system, we shall first consider an
initial system-reservoir correlated state with two-photon
correlation $\langle a(0)b_{k}(0)\rangle \neq 0$. We assume that the
cavity field is correlated initially with the first resonator mode
of the CROW in terms of a two-mode entangled squeezed vacuum state
\cite{TMSS1} as
\begin{eqnarray}
|\psi_{ab_1}(0)\rangle=\exp(-r_se^{-i\theta_s}ab_1+r_se^{i\theta_s}a^\dag
b_1^\dag)|0_a0_{b_1}\rangle,\label{cstate1}
%&=&\sec r_s\sum_{n=0}^\infty[-\exp(i\theta_s)\tanh %r_s]^n|n_an_{b_1}\rangle.\nonumber\\
\end{eqnarray}
and the other resonators in the CROW are in vacuum, with $r_s$ and
$\theta_s$ being the squeezing parameter and the reference phase,
respectively. The strength of the nonclassical correlations
(entanglement) contained in the above state increases with the
increasing of the squeezing parameter $r_s$ \cite{TMSS2}. The
reduced density matrices of the cavity field and the resonator mode
from Eq.~(\ref{cstate1}) is a mixed state which can be expressed as
\begin{eqnarray}
\rho_{a/b_1} (0)=\sum_{n=0}^\infty\frac{\sinh^{2n}r_s}
{(\sinh^2r_s+1)^{n+1}}|n_{a/b_1}\rangle\langle
n_{a/b_1}|,\label{rstate}
\end{eqnarray}
which is indeed of a single-mode thermal state with average thermal
photon number $n_{a/b_1}(0)=\sinh^2r_s$. Based on the same Fourier
transformation, it follows that the initial system-reservoir
correlations are then given by
\begin{subequations}
\label{tpc}
\begin{eqnarray}
&&\langle a(0)b_k(0)\rangle=\sqrt{\frac{1}{2\pi}}\sinh2r_se^{i\theta_s}\sin k,\\
&&\langle a(0)b_k^\dag(0)\rangle=0,
\end{eqnarray}
\end{subequations}
namely, the initial Gaussian state only has initial two-photon correlation
between the system and the reservoir.

With the initial system-reservoir correlations in Eq.~(\ref{tpc}), we obtain
from Eq.~(\ref{core}) that $\nu_1(t)=0=\upsilon_0(t)=\upsilon_2(t)$
and
\begin{subequations}
\label{core1s}
\begin{align}
&\nu_2(t)=-i\frac{\sinh2r_se^{i\theta_s}}{\sqrt{2\pi}}\mathcal{F}(t),
\\ &\upsilon_1(t)=\frac{2}{\pi}\sinh^2r_s|\mathcal{F}(t)|^2,
\end{align}
\end{subequations}
where
\begin{align}
\mathcal{F}(t)&=\int_0^td\tau\sum_kg_k\sin(k)e^{-i\omega_k\tau}u(t-\tau)\notag
\\
&=\frac{\eta}{\sqrt{2\pi}}\int_0^td\tau\int_0^\infty d\omega
\sin[k(\omega)]e^{-i\omega \tau}u(t-\tau).
\end{align}
The last line of the above equation has been applied to the
waveguide band structure given in Eq.~(\ref{strength}), so that
$\eta=\frac{\lambda}{\lambda_0}$ and
$\sin[k(\omega)]=\frac{1}{2\lambda_0}\sqrt{4\lambda_0^2-(\omega-\omega_0)^2}$
with $\omega_0-2\lambda_0\leq\omega\leq \omega_0+2\lambda_0$.

After obtaining the time-dependent functions $\nu_j(t)$ and
$\upsilon_j(t)$ given above, Eq.~(\ref{e}) becomes
\begin{subequations}
\begin{align}
&\langle a(t)\rangle =0 , \\
&n(t)=|u(t)|^2n_a(0)+\upsilon_1(t),\\
&s(t)=2u(t)\nu_2(t).\label{pjgz}
\end{align}
\end{subequations}
This solution indicates that for the given initial thermal
state $\rho_a(0)$ in Eq.(\ref{rstate}), the cavity field at time $t$ is in a squeezed
thermal state \cite{marian}, which can be expressed as
\begin{eqnarray}
\rho(t)=S_a[r(t)]\rho_{\rm th}(t) S_a^\dag[r(t)],
\end{eqnarray}
where the single-mode squeezing operator
\begin{align}
S_a[r(t)]=\exp[-\frac{r(t)}{2}e^{-i\theta(t)}a^2+\frac{r(t)}{2}e^{i\theta(t)}a^{\dag2}],
\end{align}
with the squeezing parameters
\begin{align}
r(t)=\frac{1}{4}\ln\frac{n(t)+|s(t)|+1/2}{n(t)-|s(t)|+1/2},
\end{align}
and $\theta(t)=\arg[s(t)]$. The thermal state
\begin{align}
\rho_{\rm th}(t)=\sum_{k}\frac{[\bar{n}(t)]^n}{[\bar{n}(t)+1]^{k+1}}|n_a\rangle\langle
n_a|,
\end{align}
where the average thermal photon number
$\bar{n}(t)=\sqrt{(n(t)+1/2)^2-|s(t)|^2}-1/2$. By defining the
quadrature operators $X=(a + a^\dag )/\sqrt{2}$ and $Y=(a - a^\dag
)/\sqrt{2i}$, the covariance matrix  are given by
\begin{align}
\begin{pmatrix} \Delta X^2  &  \Delta\{XY\} \\
 \Delta \{YX\}& \Delta Y^2 \end{pmatrix}
 =\Big[\bar{n}(t) + \frac{1}{2}\Big]\Big[\frac{\cosh 2r(t)}{2} I ~~\notag \\
 + \frac{\sinh 2r(t)}{2} \begin{pmatrix} \cos\theta(t) & \sin\theta(t) \\
\sin\theta(t) & -\cos\theta(t) \end{pmatrix} \Big].
\end{align}
If $\bar{n}(t)=0$, the above covariance matrix is reduced to the
standard form for a pure squeezed state \cite{Zhang90867}.
Obviously, the squeezed thermal state squeezes the thermal-state
fluctuation $\bar{n}(t) +1/2$. Thus, the squeezing in the squeezed thermal state
can be described by the squeezing parameter $r(t)$.  If
there is no initial system-reservoir correlation, then we have
$\nu_2(t)=0$ so that $s(t)=0$ which leads to the squeezing
parameter $r(t)=0$.

\begin{figure*}
\centerline{\scalebox{0.2}{\includegraphics{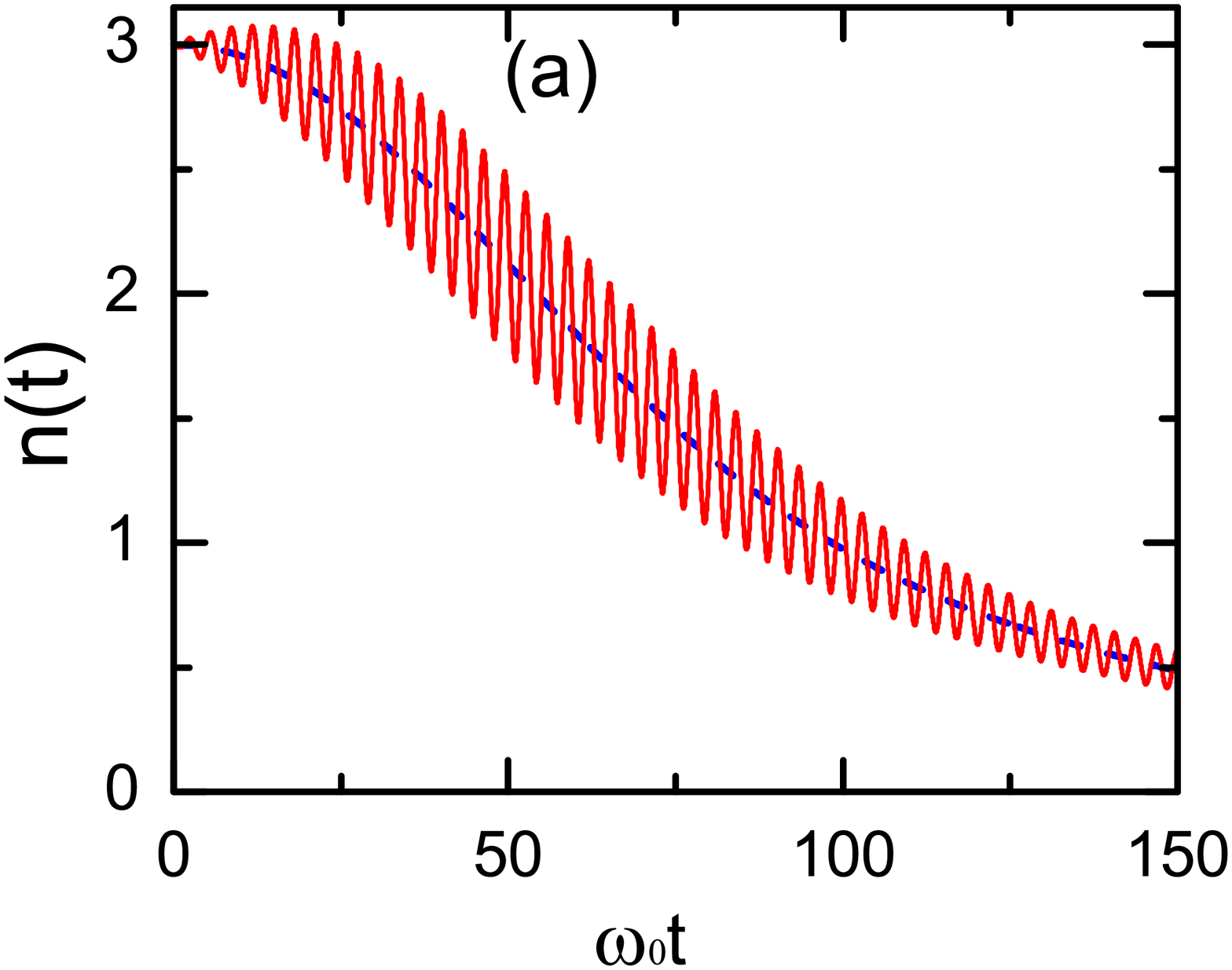}}
\scalebox{0.2}{\includegraphics{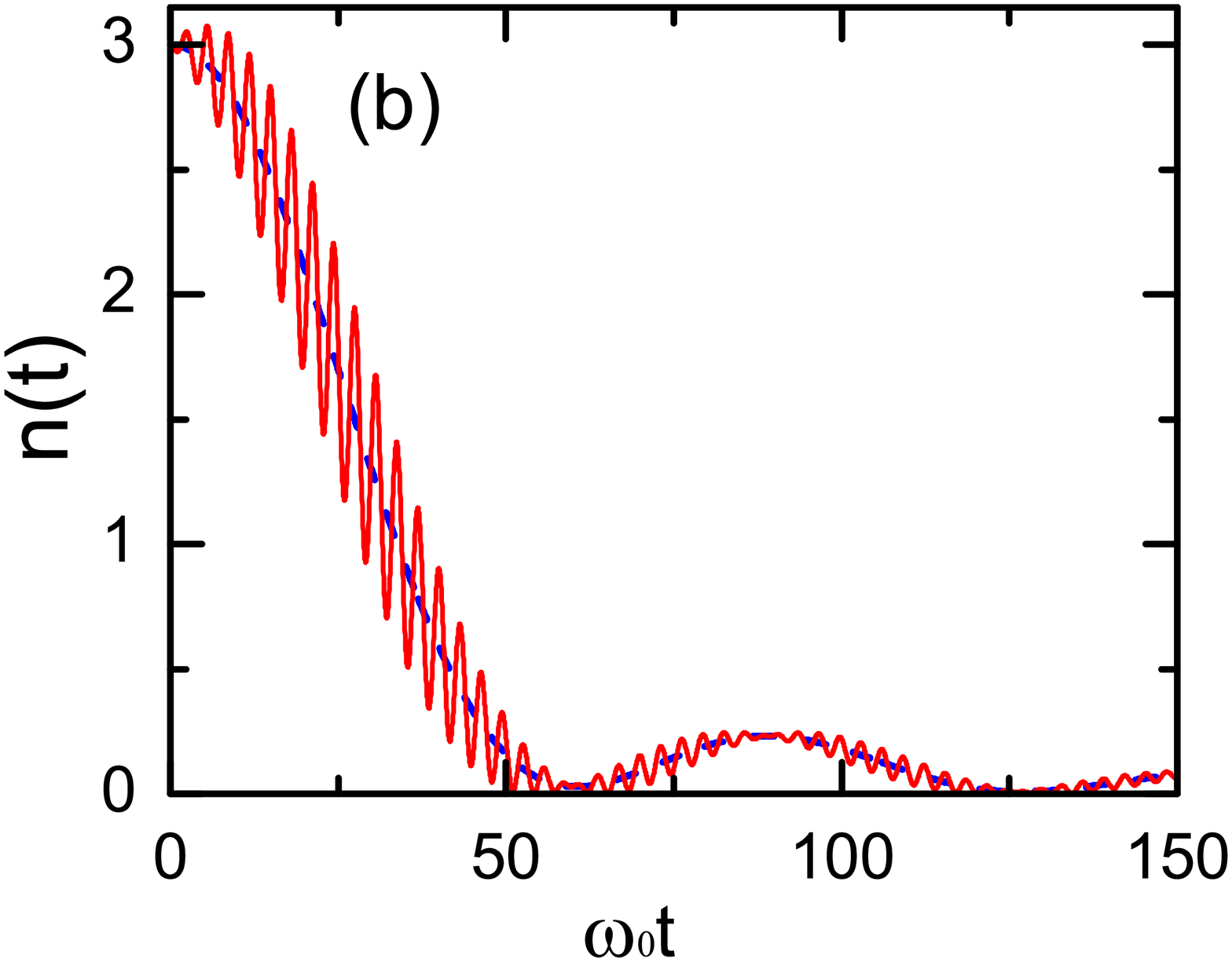}}\scalebox{0.2}{\includegraphics{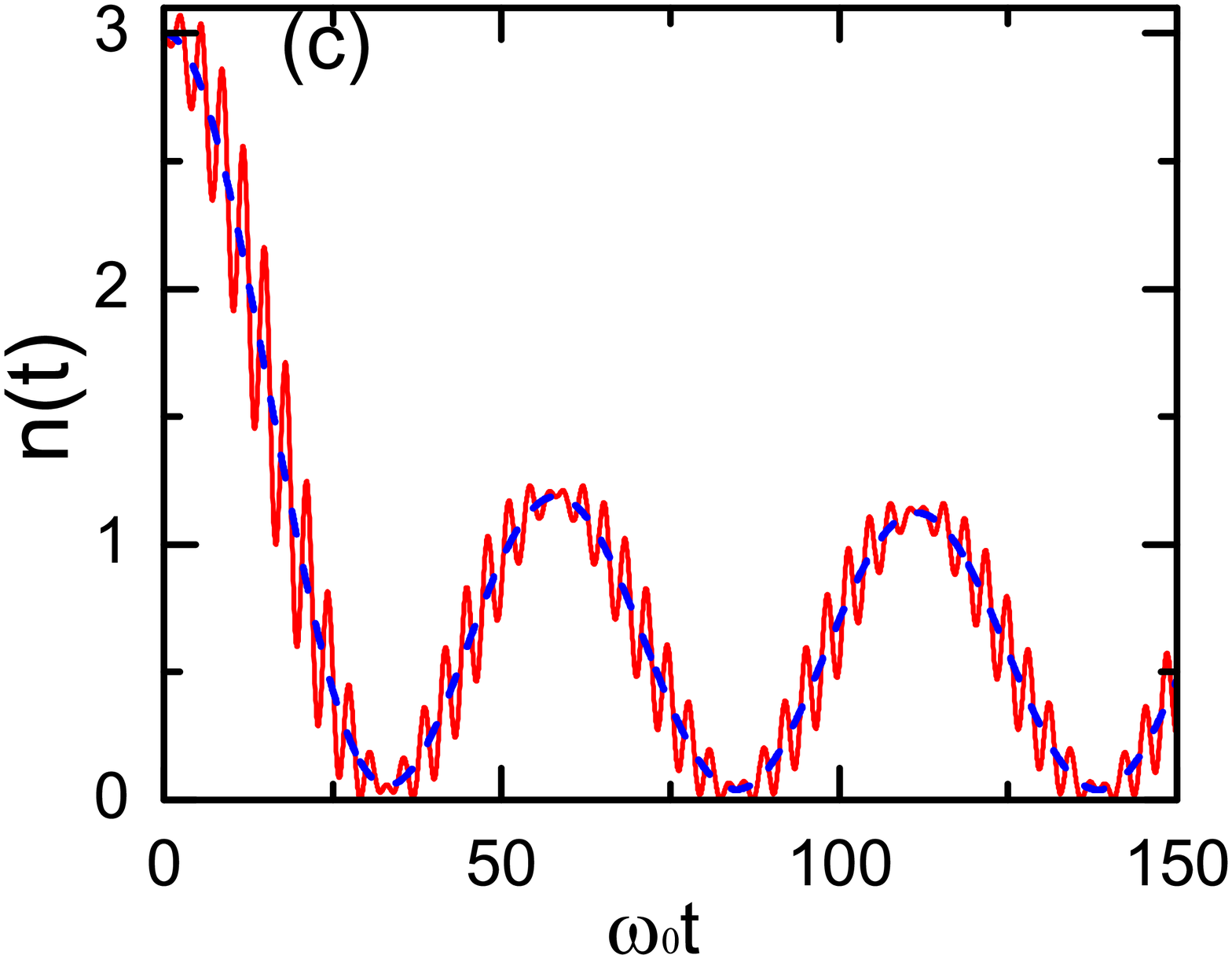}}}
\caption{The time evolution of the cavity field intensity $n(t)$
with the initial correlation $\langle a(0)b_1^\dag(0)\rangle$ (red solid
line) and without such correlation (blue dashed line) for
$\omega_c=\omega_0$, $\lambda_0=0.025\omega_0$, $\bar{n}_a=6$,
$\bar{n}_{b_1}=0$, $\vartheta=\frac{\pi}{2}$, $\eta=0.4$ in (a),
$\eta=1.2$ in (b), and $\eta=2.0$ in (c).} \label{fig3}
\end{figure*}

In Fig.~\ref{fig2}, the time evolution of the cavity intensity
$n(t)$ and the squeezing parameter $r(t)$ are plotted for the different
coupling strengths $\eta=\lambda/\lambda_0$. As shown in
Fig.~\ref{fig2}~(a), for a weak coupling ($\eta=0.4$), the cavity
intensity decays monotonically and eventually approaches to zero, as
a result of the Markovian damping dynamics at zero temperature.
Also, the small but non-zero squeezing parameter $r(t)$ indicates
that the initial two-photon correlation $\langle a(0)b_k(0)\rangle$
between the system and reservoir induces a small squeezing effect to
the cavity field in the beginning. However, the long-time behavior
of the squeezing parameter shows that the effect of the initial
system-reservoir correlation is washed out in the long-time limit,
which is also consistent with the Markovian dynamics. In contrast,
by increasing the coupling strength, as depicted in
Fig.~\ref{fig2}~(b), the cavity intensity decays rather fast in the
beginning and then it revives and damps with oscillation in which
some non-Markovian memory effect appears. Interestingly, the
squeezing parameter $r(t)$ shows a similar behavior of the
non-Markiovan effect, except for the beginning where the initial
two-photon correlation $\langle a(0)b_1(0)\rangle$ generates a
stronger squeezing effect to the cavity field, in comparison with
the weak coupling case. When the coupling strength continues
increasing to $\eta=2.0$ (the strong non-Markovian regime
\cite{Wu1018407}), the cavity intensity decays faster in the very
beginning and then revives and keeps oscillating without damping
from then on, see Fig.~\ref{fig2}~(c). In this situation, we find
that the initial-correlation-induced squeezing dynamics also
oscillates over all the time. Therefore, we can conclude that the
initial two-photon correlations $\langle a(0)b_k(0)\rangle$ can lead
to a nontrivial squeezing dynamics of the cavity field,
as a consequence of strong non-Markovian memory dynamics, but it is
negligible in the steady-state limit in the Markovian regime.

\subsection{Initially system-reservoir correlated mixed thermal states}
Next, we investigate the effect of the initially system-reservoir
correlated state with the correlation $\langle
a(0)b_k^\dag(0)\rangle \neq 0$. To this end, we consider an
initially mixed state
\begin{eqnarray}
\rho_{ab_1}(0)=B(\vartheta)\rho_a\otimes\rho_{b_1}B^\dag(\vartheta),
\label{ics}
\end{eqnarray}
where the operator
$B(\vartheta)=\exp[\frac{\vartheta}{2}(ab_1^\dag-a^\dag b_1)]$ and
the density operators $\rho_{a/b_1}$ represent the thermal states
with average thermal photon numbers $\bar{n}_{a/b_1}$. This
initially correlated state can be formed via the bilinear coupling
between the cavity field $a$ and the resonator mode $b_1$ in the
thermal states, and note that nonclassical entanglement are not
present in this initially correlated state \cite{kim}. A direct
calculation shows that the initial system-reservoir correlations
\begin{subequations}
\begin{align}
&\langle a(0)b_k(0)\rangle=0, \\
& \langle
a(0)b_k^\dag(0)\rangle=\frac{\sin\vartheta}{\sqrt{2\pi}}(\bar{n}_a-\bar{n}_{b_1})\sin
k.
\end{align}
\end{subequations}
For the initially correlated state of Eq.(\ref{ics}), it is easy to
find that the reduced density matrices  $\rho_a(0)$ and
$\rho_{b_1}(0)$ of the cavity field and the resonator mode $b_1$ are
also the thermal states with the average thermal photon numbers
$n_{a}(0)=\frac{1}{2}[\bar{n}_a+\bar{n}_{b_1}+(\bar{n}_a-\bar{n}_{b_1})\cos\vartheta]$
and
$n_{b_1}(0)=\frac{1}{2}[\bar{n}_a+\bar{n}_{b_1}-(\bar{n}_a-\bar{n}_{b_1})\cos\vartheta]$,
respectively. From Eq.(\ref{core}), we obtain
\begin{eqnarray}
\nu_1(t)&=&-i
\frac{(\bar{n}_a-\bar{n}_{b_1})\sin \vartheta}{\sqrt{2\pi}}\mathcal{F}(t),\\
\upsilon_1(t)&=&\frac{2n_{b_1}(0)}{\pi}|\mathcal{F}(t)|^2,
\end{eqnarray}
and $\nu_2(t)=0$, and $\upsilon_0(t)=0$, $\upsilon_2(t)=0$. Thus,
the corresponding physical observables of the cavity system for the
above initially correlated state of Eq.(\ref{ics}) are given by
\begin{eqnarray}
 n(t)
=|u(t)|^2n_a(0)+2Re[u^*(t)\nu_1(t)]+\upsilon_1(t),
\end{eqnarray}
and $\langle a(t)\rangle=0$, $ s(t) =0$. It indicates that the
cavity state remains in a thermal state over all the time with the
cavity field intensity $\sim n(t)$.

In Fig.~\ref{fig3}, the time evolution of the cavity field intensity
$n(t)$ is plotted for different coupling strengths $\eta$ between
the nanocavity and the waveguide. Fig.~\ref{fig3}~(a) shows the the
average photon number for a weak coupling $\eta=0.4$. It reveals
that the intensity of the cavity field decays monotonically to a
steady-state value, as a character of the Markovian dynamics. The
initial system-reservoir correlation $\langle
a(0)b_k^\dag(0)\rangle$ leads to the intensity oscillating around
the decay line of the case of the initially uncorrelated state. The
amplitude of the local oscillations increases in the beginning and
then decreases to a unnoticeable effect as time develops. In other
words, the effect of the initial correlation $\langle
a(0)b_k^\dag(0)\rangle$ will be washed out in the steady limit. With
the increasing of the coupling strength, the intensity no longer
monotonically decays and some revival phenomena occur as a character
of the non-Markovian memory dynamics \cite{Wu1018407}, as depicted
in Fig.~\ref{fig3}~(b). When the coupling is increased to $\eta=2.0$
as a strong coupling value, we see from Fig.~\ref{fig3}~(c) that the
intensity and the initial system-reservoir induced oscillation keep
oscillating in the whole time regime. In other words, the effect
resulted from the initial system-reservoir correlation in the
non-Markovian regime will not be washed out by the interaction
between the system and the reservoir.

\section{Summary}
In summary, we investigate the dynamics of open quantum systems in
the presence of initial system-reservoir correlations. We take the
photonic cavity system coupled to a non-Markovian reservoir as a
specific open quantum system. By solving the exact dynamics of the
cavity system, the effects of the initial correlations are
explicitly built into the solution of the cavity field intensity and
the two-photon correlation function. We also derive the
time-convolutionless exact master equation which incorporates with
the initial system-reservoir correlations. The non-Markovian memory
effects are fully embedded into the time-dependent coefficients in
the master equation. The fluctuation coefficient $\gamma_2(t)$ is
modified by the initial system-reservoir photonic scattering
correlation but the frequency shift $\Delta (t)$ of the cavity and
the dissipation coefficient $\gamma_1(t)$ remain unchanged. However,
the initial two-photon correlation between the system and the
reservoir induces two-photon decoherence terms in the master
equation, which can lead to photon squeezing in the cavity. We also
take a nanocavity coupled to a coupled resonator optical waveguide
(serving as a structured reservoir) as an experimentally realizable
system, from which we find that the effects of the initial
correlations are fragile for a Markovian reservoir but play an
important role in the non-Markovian regime. In fact, in the strong
non-Markovian regime, the initial two-photon correlation between the
cavity and the reservoir can induce oscillating squeezing dynamics
in the cavity. But in Markovian regime, the effects of the initial
system-reservoir correlations will be washed out in the steady-state
limit.

\section*{Acknowledgment}
This work is supported by the National Science Council of ROC under
Contract No. NSC-99-2112-M-006-008-MY3, the National Center for
Theoretical Science of Taiwan, National Natural Science Foundation
of China (Grant No.10804035), and SDRF of CCNU (Grant No. CCNU
09A01023).

\end{document}